\documentclass[12pt,preprint]{aastex}
\begin{document}
\title{Reconstructing the Triaxial Shapes of Dark Matter Halos from the 
Anisotropic Spatial Distributions of their Substructures in the Concordance 
Cosmology} 
\author{\sc Jounghun Lee\altaffilmark{1} and Xi Kang\altaffilmark{2,3}}
\altaffiltext{1}{Astronomy Program, School of Earth and Environmental 
Sciences, Seoul National University, Seoul 151-742 , Korea: 
jounghun@astro.snu.ac.kr}
\altaffiltext{2}{Shanghai Astronomical Observatory; the Partner Group of 
MPA, Nandan Road 80, Shanghai 200030, China}
\altaffiltext{3}{Astrophysics Department, University of Oxford, Oxford, 
OX1 3RH, UK}
\begin{abstract} 
We develop an algorithm to reconstruct the triaxial shapes of dark 
matter halos from the anisotropic spatial distributions of their 
substructures for the concordance background cosmology. 
First, we construct an analytic model for the anisotropic spatial 
distribution of dark halo substructures under the assumption that the 
tidal field with non-zero trace in the triaxial mass distribution of the 
host halos generates the substructure bulk motions toward the major principal 
axes of the hosts. Our analytic model implies that the degree of anisotropy 
depends sensitively on the triaxiality of the host halos as well as the 
correlation between the substructure locations and the tidal shear field. 
Second, we set the axis ratios of the triaxial host halos as free parameters 
in the analytic model, fit the model to the numerical results from 
high-resolution N-body simulation of the concordance cosmology, and 
reconstruct the two axis-ratios of the host halos from the best-fit values 
of the free parameters. The comparison of the reconstructed axis ratios with 
the numerical results reveals a good agreement. Finally, we conclude that our 
analytic model may provide a physical understanding of the anisotropic spatial 
distribution of dark halo substructures and a new way to reconstruct in 
principle the triaxial shapes of dark matter halos from the observables.
\end{abstract} 

\section{INTRODUCTION}

The phenomenon that the satellite galaxies exhibit strong degree of anisotropy 
in their spatial distributions has recently provoked two-fold controversy:
(i) whether the anisotropy is along the minor or major axes of the host 
galaxies; (ii) whether it is consistent with the cold dark 
matter (CDM) paradigm or not.

In fact, the first observational report of this phenomenon dates back to 
more than three decades ago. \citet{hol69} analyzed the two dimensional 
projected positions of the satellite galaxies of nearby spirals, and found 
an effect (the Holmberg effect) that the satellite locations within the 
projected radii $\le 50$ kpc are quite biased toward the regions around the 
minor axes of their hosts.  In the beginning, the Holmberg effect was 
suspected not to be real, but due to either the poor number statistics or 
the projection effects \citep{haw-pee75,sha-etal79,mac-etal82}.  It is only 
very recently when this effect gains supporting evidence for its existence.  
\citet{sal-lam04} used a large sample of spiral galaxies and their satellites 
collected from the Two Degree Field Galaxy Redshift Survey 
\citep[][2dFGRS]{col-etal01}, and found that the satellites have indeed 
a strong tendency to be located near the minor axes of their hosts as 
Holmberg claimed more than three decades ago.

In spite of the compelling evidence presented in \citet{sal-lam04}, the 
Holmberg effect has yet to be confirmed, because another recent report 
arouse a new suspicion. \citet{bra05} conducted an analysis 
of the satellite galaxies about isolated host galaxies in the second data 
release of the Sloan Digital Sky Survey \citep[][SDSS]{str-etal02}, 
and found that the anisotropy of the satellite spatial distribution is along 
not the minor but the {\it major} axis of the host galaxies, in direct
contrast with the Holmberg effect. 

A cosmological turmoil surrounding the satellite anisotropy was caused by 
the other very recent report. Kroupa, Theis, \&  Boily (2005) showed that 
the Milky Way (MW) dwarf satellites are located near polar plane 
\citep{lyn82,maj94}, and claimed that this anisotropic distribution of MW 
satellites is inconsistent with being drawn from a CDM substructure 
population. Shortly after \citet{kro-etal05}, a flurry of numerical studies 
demonstrated that the CDM model indeed predicts anisotropic satellite 
distributions.  Kang et al. (2005) showed by their high-resolution N-body 
simulation that the satellites are likely to lie in the plane defined by 
the major and intermediate axes of the host halo. 
\citet{zen-etal05} confirmed the 
findings of \citet{kang-etal05} by providing the numerical results that 
the dark halo substructures on galactic scales are preferentially 
located along the major axes of the triaxial mass distributions of their 
hosts, proving that the CDM paradigm does not predict the isotropic spatial 
distribution of satellite galaxies \citep[see also,][]{kne-etal04}. 
\citet{lib-etal05} also performed high-resolution simulations of the 
galactic CDM halos to verify  that the satellite anisotropy is not 
inconsistent with the CDM cosmology.

Yet, an important problem highlighted by \citet{kro-etal05} still remains 
a mystery: why are those halos forming the luminous dSphs  located in a 
disk-like configuration almost perpendicular to its host galaxy ?  
This question in fact holds a crucial key to understanding the formation 
of galaxies since if dSphs are assumed to follow the dark matter distribution, 
the observed polar alignment of dSphs indicates a disagreement between 
the baryons of the host galaxy and its dark matter halos. 
 
To address this issue and confirm the Holmberg effect, it is necessary first  
to understand the physical  origin of the satellite anisotropy. Now that 
several numerical studies proved that the satellite anisotropy is consistent 
with the CDM model, it should be possible to construct a theoretical model 
for the anisotropic spatial distribution of satellites in the CDM context. 
The standard scenario explains that the anisotropy in the satellite 
distribution is originated from the anisotropic infall of material along 
filaments (West 1989, Bond, Kofman, \& Pogosyan 1996). 
But, their explanations were inconclusive, making no quantitative prediction 
for the degree of anisotropy expected on different scales. 
 
Very recently, Lee, Kang, \& Jing (2005, hereafter LKJ05) pointed out 
the incompleteness of the anisotropic infall scenario in their studies of  
the anisotropic {\it orientations} (not the anisotropic locations) 
of dark halo substructures.  They claimed that since the anisotropic infall 
is a primordial effect, being quickly damped away by the subsequent nonlinear 
process like the secondary infall, it is unlikely to be the direct cause 
for the anisotropic orientations of substructures.  
They assumed that the substructure anisotropic orientations must be caused 
by the tidal interactions with the gravitational fields in the triaxial mass 
distribution of their host halos, and constructed an analytic model for the 
alignments between the major axes of the substructures and their host halos 
from the physical principles, which turned out to be in good agreement with 
simulation results.

Motivated by the success of LKJ05 model, here we attempt to construct a 
similar analytic model for the {\it anisotropic spatial distribution} of 
dark halo substructures assuming that the anisotropy in the spatial 
locations of substructures  are also induced by the tidal interaction 
with the gravitational field in the triaxial mass distribution of host halos. 
Our final goal is to develop an algorithm for reconstructing the triaxial 
shapes of host halos from the observable anisotropic spatial distribution 
of their substructures by using this analytic model.
Throughout this paper, we assume a {\it concordance} $\Lambda$CDM cosmology 
with $\Omega_{0}=0.3$, $\Omega_{\Lambda,0}=0.7$, and $h=0.7$.

\section{PHYSICAL ANALYSIS}

\citet{lee-pen01} claimed that the position vectors of dark halos 
are related to the tidal shear tensors, and suggested that the 
relationship between the two quantities can be written as a first-order 
correlation (see eq. [12] in Lee \& Pen 2001).  They supported their claims 
by N-body simulations \citep[see also][]{lee-pen00} where it was found 
that the tidal shear tensors are indeed correlated with the halo position 
vectors at first order. Although they did not show that it also holds on 
the substructure scale, we extrapolate here their claim to the 
substructure scale, and adopt an {\it Ansatz} that the position 
vectors of dark halo substructures are correlated with the tidal shear 
tensors of their host halos at first order.  

Suppose that a subhalo lies in the perturbation potential field 
$\Psi$ inside its triaxial host halo. The potential $\Psi$ is assumed to 
be smoothed on the host halo mass scale. Let ${\bf q} = (q_i)$ be 
the initial position of the subhalo when there is no interaction with the 
host halo potential field, and let also ${\bf x} = (x_i)$ represent the 
final position of the subhalo under the interaction of the host halo 
potential. Both vectors, ${\bf x}$ and ${\bf q}$, are defined in the 
center of mass frame of the host halo.  Following the linear scheme 
\citep{zel70}, the relation between $(x_i)$ and $(q_i)$ in the host 
halo potential may be written as 
\begin{equation}
\label{eqn:zel}
x_{i} = -D_{+}\partial_{i}\Psi({\bf q}),
\end{equation}
where $D_{+}$ is the linear growth rate of the density field. Note that 
${\bf x}$ in equation (\ref{eqn:zel}) represents the displacement of the 
subhalo position generated by the interaction with their host halo 
gravitational field. 

Taylor-expanding $\Psi$ to second order around the host halo center of mass, 
we have  
\begin{equation}
\label{eqn:psi}
\Psi = \Psi(0) + q_{k}\partial_{j}\Psi|_{0} + 
\frac{1}{2}q_{j}q_{k}\partial_{j}\partial_{k}\Psi|_{0}.
\end{equation}
Differentiating equation (\ref{eqn:psi}) gives
\begin{equation}
\label{eqn:difpsi}
\partial_{i}\Psi = \partial_{i}\Psi|_{0} + 
q_{k}\partial_{i}\partial_{k}\Psi|_{0}.
\end{equation}
Putting equation (\ref{eqn:difpsi}) into equation (\ref{eqn:zel}), 
we end up with 
\begin{equation}
\label{eqn:xq}
x_{i} \propto q_{k}T_{ik}(0), 
\end{equation}
where ${\bf T} = (T_{ik})$ is the tidal shear tensor defined as
$\partial_{i}\partial_{k}\Psi$. Equation (\ref{eqn:xq}) implies that 
the subhalo displacement is a bulk motion generated by the nonzero 
quadrupole moment of the triaxial gravitational potential of the host halo. 

One might think that since equation (\ref{eqn:zel}) is derived from the 
linear perturbation theory based on the Zel'dovich approximation, it should 
be invalid in the highly nonlinear regime. But, \citet{lee-pen01} showed 
by N-body simulations that even in the highly nonlinear regime the 
correlation between the halo position vector and the tidal shear tensor 
can be written as first order like equation (\ref{eqn:xq})

To investigate how the direction of ${\bf x}$ 
is distributed given ${\bf T}$, we first consider the covariance 
$\langle x_{i}x_{j} | {\bf T} \rangle$.   Using equation (\ref{eqn:xq}), 
we have $\langle x_{i}x_{j} | {\bf T} \rangle 
\propto \langle q_{k}q_{l}T_{ik}T_{lj}\rangle$. 
If the two tensors, $(q_{i}q_{j})$ and $(T_{ik}T_{kj})$, were mutually 
independent, then, we would end up with having a simple expression,  
$\langle q_{k}q_{l}T_{ik}T_{jl}\rangle =  
\langle q_{k}q_{l}\rangle T_{ik}T_{lj} \propto T_{ik}T_{kj}$ as 
$\langle q_{k}q_{l}\rangle \propto \delta_{kl}$. However, it it not 
necessarily true that $(q_{i}q_{j})$ is uncorrelated with $(T_{ik}T_{kj})$.

To take into consideration the possible correlation between $(q_{i}q_{j})$ 
and $(T_{ik}T_{kj})$, and to overcome the limited validity of the linear  
approximation, we use the following parametrized formula as 
\citet{lee-pen01} did:
\begin{equation}
\label{eqn:quadratic}
\langle {x}_{i}{x}_{j} | {\bf T} \rangle =
\frac{1 - s}{3}\delta_{ij} + s\check{T}_{ik}\check{T}_{kj},
\end{equation}
where $x_{i}$ is the rescaled position vector (rescaled in a sense that it 
does not have a dimension of ``length''), $\check{\bf T} = (\check{T}_{kj})$ 
is the unit tidal shear tensor with {\it non-zero trace}: 
$\check{T}_{ik} \equiv T_{ik}/\vert{\bf T}\vert$ and 
${\rm Tr}(\check{\bf T})\ne 0$  (we use the notation of $\check{\bf T}$ to 
distinguish it from the traceless unit tidal shear tensor $\hat{\bf T})$, 
and $s$ is a free parameter in the range of $[-1,1]$ to quantify the 
correlation between $({x}_{i}{x}_{j})$ and $(\check{T}_{ik}\check{T}_{kj})$. 
In the linear theory, the value of $s$ is given as unity. 

Note that equation (\ref{eqn:quadratic}), although it is derived from the 
linear perturbation theory, has already included the nonlinear effect. 
The correlation parameter, $s$, quantifies how the correlation between 
${\bf x}$ and ${\bf T}$ differs from the linear theory prediction. 
The degree of the deviation of its value from unity measures how strong 
the nonlinear effect is.

It is also worth noting that equation (\ref{eqn:quadratic}) bears a close 
resemblance to the quadratic formula used by LKJ05 for the correlation 
between the dark halo angular momentum ${\bf L} = (L_{i})$
and the unit traceless tidal shear field $\hat{\bf T} = (\hat{T}_{ij})$ 
defined as  $\hat{T}_{ij} \equiv \tilde{T}_{ij}/\vert\tilde{\bf T}\vert$ 
with $\tilde{T}_{ij} \equiv T_{ij} - {\rm Tr}({\bf T})\delta_{ij}/3$
\begin{equation}
\label{eqn:spin-shear}
\langle L_{i} L_{j} | {\bf T} \rangle =
\frac{1 + c}{3}\delta_{ij} - c\hat{T}_{ik}\hat{T}_{kj},
\end{equation}
where the correlation parameter $c$ is in the range of $[0,1]$. LKJ05 
used equation (\ref{eqn:spin-shear}) to investigate the alignments between 
the major axes of substructures and their host halos.

Similar as equations (\ref{eqn:quadratic}) and  (\ref{eqn:spin-shear}) 
look, it is very important to recognize the key differences between the two. 
In equation (\ref{eqn:spin-shear}), $\hat{\bf T}$ is traceless. Consequently, 
${\bf L}$ is preferentially aligned only with the {\it intermediate} principal 
axis of ${\bf T}$. Whereas in equation (\ref{eqn:quadratic}) $\check{\bf T}$ 
has non-zero trace. Consequently,  ${\bf x}$ is preferentially aligned 
with either the {\it minor} or the {\it major} principal axis of ${\bf T}$:  
If $s$ is negative, then ${\bf x}$ is aligned with the minor axis of ${\bf T}$.
If $s$ is positive, ${\bf x}$ is aligned with the major axis of ${\bf T}$. 

Strictly speaking, equation (\ref{eqn:quadratic}) holds only if 
$\check{\bf T}$ and ${\bf x}$ are smoothed on the same scale. 
Here, they don't: $\check{\bf T}$ is supposed to be smoothed on the 
host halo mass scale, while ${\bf x}$ on the substructure radii. 
For simplicity, here, we just assume that  equation (\ref{eqn:quadratic}) 
still holds, ignoring the difference between the host halo and their 
substructure smoothing scales.

In a similar manner with LKJ05, assuming that the probability 
density distribution of $p({\bf x}|{\bf T})$ is Gaussian, and expressing the 
position vectors ${\bf x}$ in terms of the spherical polar coordinates in 
the principal axis frame of ${\bf T}$ as 
${\bf x} = (x\sin\alpha\cos\beta,x\sin\alpha\sin\beta,x\cos\alpha)$
where $x \equiv |{\bf x}|$, $\alpha$ and $\beta$: the polar and the azimuthal 
angles of ${\bf x}$, respectively,  we derive the probability density 
distribution of the cosines of the polar angles, $p(\cos\alpha)$ as:
\begin{eqnarray}
p(\cos\alpha) &=& \frac{1}{2\pi}\prod_{i=1}^{3}
\left(1-s+3s\check{\lambda}^{2}_{i}\right)^{-\frac{1}{2}}\times \nonumber \\
&&\int_{0}^{2\pi}
\left(\frac{\sin^{2}\alpha\cos^{2}\beta}{1-s+3s\check{\lambda}^{2}_{1}} + 
\frac{\sin^{2}\alpha\sin^{2}\beta}{1-s+3s\check{\lambda}^{2}_{2}} + 
\frac{\cos^{2}\alpha}{1-s+3s\check{\lambda}^{2}_{3}}\right)^{-\frac{3}{2}}
d\beta.
\label{eqn:alpha_dis}
\end{eqnarray}
\begin{equation}
\label{eqn:hatlam_lam}
\check{\lambda}^{2}_{1}+\check{\lambda}^{2}_{2}+\check{\lambda}^{2}_{3}= 1, 
\qquad \check{\lambda}_{1}+\check{\lambda}_{2}+ \check{\lambda}_{3}=
\frac{\lambda_{1} + \lambda_{2} + \lambda_{3}}
{\left(\lambda^{2}_{1}+\lambda^{2}_{2}+\lambda^{2}_{3}\right)^{1/2}},
\end{equation}
where $\{\lambda_i\}_{i=0}^{3}$ are the three eigenvalues of the tidal shear 
tensor ${\bf T}$ in a decreasing order, and $\{\check{\lambda_i}\}_{i=1}^{3}$ 
are the unit eigenvalues of $\check{\bf T}$ defined as 
$\check{\lambda_i} \equiv \lambda_{i}/
\left(\lambda^{2}_{1}+\lambda^{2}_{2}+\lambda^{2}_{3}\right)^{1/2}$. 

In the linear density field, the sum of all the three eigenvalues of the 
tidal shear tensor equals the linear density contrast: 
$\lambda_{1}+\lambda_{2}+\lambda_{3}=\delta$. 
According to the spherical dynamical collapse model, a gravitational 
bound object forms when the linear density contrast reaches its critical 
value, $\delta_{c}$. Hence, one can say that inside the host halo
$\lambda_{1}+\lambda_{2}+\lambda_{3}=\delta_{c}$. 
The value of the critical density contrast, $\delta_{c}$, scales with the 
redshift $z$ as $\delta_{c}(z) = D_{+}(z)\delta_{0c}$ where $\delta_{c0}$ 
is the critical value at $z=0$, which depends very weakly on cosmology 
\citep{eke-etal96}. For example, $\delta_{c0}=1.686$ for the flat universe 
of closure density. Now, one can rewrite equation (\ref{eqn:hatlam_lam}) as 
\begin{equation}
\label{eqn:hatlam_delc}
\check{\lambda}_{1}+\check{\lambda}_{2}+ \check{\lambda}_{3}=
\frac{\delta_{c}}
{\left(\lambda^{2}_{1}+\lambda^{2}_{2}+\lambda^{2}_{3}\right)^{1/2}}
\end{equation}
 
As LKJ05 did, we assume further that the tidal shear and the inertia 
shape tensors of the host halos are strongly correlated with each other 
in an opposite direction \citep{lee-pen00,por-etal02}, 
so that the minor and the major axes of the tidal shear tensor are the 
major and minor axes of the inertia shape tensor. 
Under this assumption,  equation (\ref{eqn:alpha_dis}) basically 
represents the probability density distribution of the cosines of the 
angles between the position vectors of the substructures and the major 
principal axes of their host halo.
Equation (\ref{eqn:alpha_dis}) reveals that $p(\cos\alpha)$ depends 
sensitively on the value of $s$: If $-1\le s < 0$, ${\bf x}$ is aligned 
with the major axis of the host halo (the minor axis of ${\bf T}$); 
If $0< s \le 1$, ${\bf x}$ is aligned with the minor axis of the 
host halo (the major axis of the ${\bf T}$); If $s=0$, there is no 
alignment between ${\bf x}$ and any principal axis of the host halo.

Note that even when $s \ne 0$, it is possible to have no alignment 
between ${\bf x}$ and the host halo axes. The distribution $p(\cos\alpha)$ 
depends not only on the value of $s$ but also on the unit eigenvalues with 
non-zero trace,  $\{\check{\lambda_i}\}_{i=0}^{3}$.  
For the case of the {\it traceless} eigenvalues of 
$\{\hat{\lambda_{i}}\}_{i=0}^{3}$ which satisfies the two constraints: 
$\sum_{i=0}^{3}\hat{\lambda_i}=0$ and $\sum_{i=0}^{3}\hat{\lambda^{2}_i}=1$ 
(see $\S 2$ in LKJ05), their values are well approximated as the average 
$\hat{\lambda_1}=-\hat{\lambda_3} = 1/\sqrt{2}$  and $\hat{\lambda_2}=0$ 
regardless of the original values of $\{\lambda_{i}\}_{i=1}^{3}$. 
However, for  $\{\check{\lambda_i}\}_{i=0}^{3}$, the constraints 
(eq.[\ref{eqn:hatlam_lam}]) are expressed in terms of the original values 
of $\{\lambda_i\}_{i=0}^{3}$. Therefore, even when the correlation parameter 
$s$ has the maximum absolute value (i.e., $\vert s \vert$ = 1), if the host 
halo has almost spherical shape satisfying $\lambda_{1}=\lambda_{2}=
\lambda_{3}=0$, then $p(\cos\alpha)$ becomes uniform, implying no anisotropy 
in the spatial distribution of substructures.

\section{HALO SHAPE RECONSTRUCTION}

To find $p(\cos\alpha)$ numerically, we use the same numerical data 
as used in LKJ05 which were generated by P$^{3}$M N-body simulations 
with $256^{3}$ particles in a $100h^{-1}$Mpc cube \citep{jin-sut98,jin-sut00} 
for the $\Lambda$CDM cosmology with $\Omega_{0}=0.3$, 
$\Omega_{\Lambda,0}=0.7$, and $h=0.7$. In the simulation data, we selected 
four dark halos on mass scales around $5-10 \times 10^{14}h^{-1}M_{\odot}$, 
and identified the self-bound substructures in each host halo by using the 
{\tt SUBFIND} routine of \citet{spr-etal01}.  For the detailed descriptions 
of the N-body simulations and the subhalo identification procedure,  
see $\S 3$ in LKJ05. 

For each host halo selected in the numerical data, we compute the 
inertia shape tensor $I_{ij} \equiv \Sigma_{\alpha} 
m_{\alpha}r_{\alpha,i}r_{\alpha,j}$ and determine the major principal axis 
by finding the eigenvector $\hat{\bf e}_{h}$ of $(I_{ij})$ corresponding to 
the largest eigenvalue.  In each host halo principal axis frame, 
we measure the unit position vectors of the subhalos $\hat{\bf x}_{s}$, and 
the cosines of the angles, $\alpha$, between the major axis 
of the host halo and the locations of their subhalos by computing 
$\cos\alpha \equiv\vert \hat{\bf e}_{h}\cdot \hat{\bf x}_{s}\vert$.  
We find the probability density distribution, $p(\cos\alpha)$ 
by counting the subhalo number density as a function of $\cos\alpha$. 

We fit our analytic model (eq.[\ref{eqn:alpha_dis}]) to this numerically 
determined $p(\cos\alpha)$ with setting the three parameters, 
$\check{\lambda}_{1},\check{\lambda}_{2}$ ($\check{\lambda}_{3}$ is not 
independent according to eq.[\ref{eqn:hatlam_lam}]) and $s$ as free, 
and determine the best-fit values of the free parameters by means of the 
$\chi^{2}$ statistics.  We perform this procedure for each individual host 
halo (CL\#1, CL\#2, CL\#3, CL\#4) separately at four different redshifts 
($z=0,0.5,1,1.5$). 

\begin{figure}
\begin{center}
\plotone{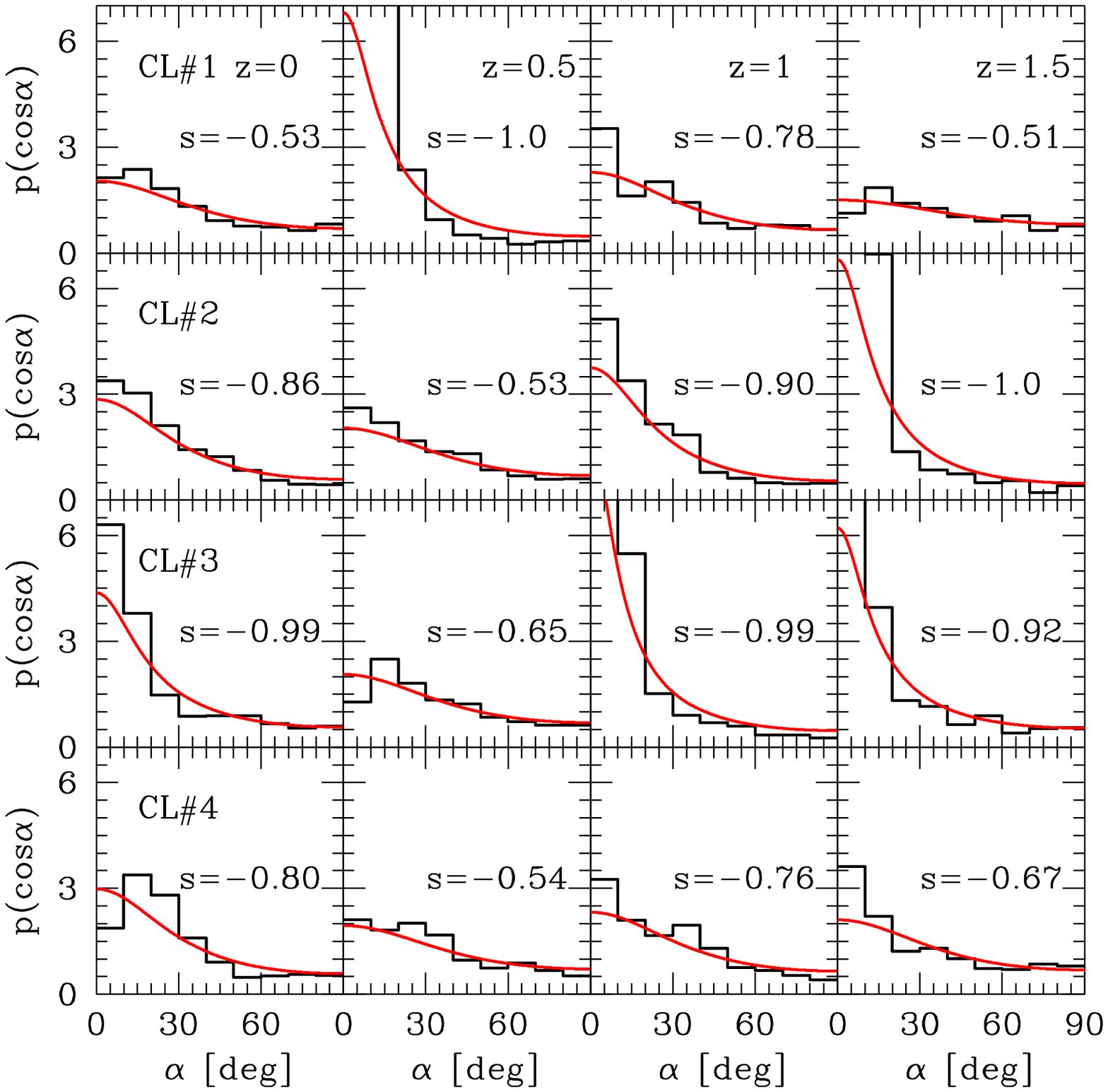}
\caption{Probability density distributions of the angles between the 
major axes of four individual host halos and the position vectors of their 
substructures at four different epochs; $z=0,0.5,1$ and $1.5$. In each panel, 
the histogram represents the simulation results, and the (red) solid 
line represent the theoretical prediction (\ref{eqn:alpha_dis}).
\label{fig:pro_dis}}
\end{center}
\end{figure}

Figure \ref{fig:pro_dis} plots the final fitting results.
In each panel, the histogram represents the numerically measured 
$p(\cos\alpha)$, while the (red) solid curve corresponds to the analytic 
distribution (eq.[\ref{eqn:alpha_dis}]) with the best-fit correlation 
parameter, $s$.
As one can see, for all cases, the best-fit correlation parameter $s$ 
has a negative value. In other words, the dark halo substructures are 
preferentially located near the {\it major} axes of their host halos, 
consistent with previous numerical findings 
\citep{kne-etal04,zen-etal05,lib-etal05}. Indeed, the analytic 
distributions agree with the numerical results for all four clusters 
at all redshifts quite well. 

As a final step, we reconstruct the triaxial shapes of the four host halos 
from their distributions $p(\cos\alpha)$. For this purpose, we adopt the 
following formula suggested recently by \citet{lee-etal05}: 
\begin{equation}
\label{eqn:axis_lambda}
\frac{a_2}{a_1} = \left(\frac{1-D_+\lambda_2}
{1-D_+\lambda_3}\right)^{1/2},\qquad 
\frac{a_3}{a_1} = \left(\frac{1-D_+\lambda_1}
{1-D_+\lambda_3}\right)^{1/2}
\end{equation}
where $a_{1},a_{2},a_{3}$ represent the three axis lengths of the triaxial 
host halos in a decreasing order. Equation (\ref{eqn:axis_lambda}) basically 
links the two axes ratios of the triaxial host halos to the eigenvalues 
of the tidal shear tensors. According to this equation, once 
$\{\lambda_i\}^{3}_{i=3}$ is determined, then one can reconstruct 
the axis-ratios of the host halos. 

Equations (\ref{eqn:hatlam_lam}) and (\ref{eqn:hatlam_delc}) allow one 
to determine the values of $\{\lambda_i\}^{3}_{i=3}$ from the best-fit 
values of $\{\check{\lambda_i}\}^{3}_{i=3}$ as  
\begin{equation}
\label{eqn:lam_hatdelc}
\lambda_{1} = \frac{\delta_c\check{\lambda_1}}{\check{\lambda_1}+
\check{\lambda_2}+\check{\lambda_3}},\quad
\lambda_{2} = \frac{\delta_c\check{\lambda_2}}{\check{\lambda_1}+
\check{\lambda_2}+\check{\lambda_3}}, \quad
\lambda_{3} = \frac{\delta_c\check{\lambda_3}}{\check{\lambda_1}+
\check{\lambda_2}+\check{\lambda_3}}
\end{equation}
We use $\delta_c = 1.686$ as an approximation and the best-fit values 
of $\check{\lambda_{1}},\check{\lambda_{2}},\check{\lambda_{3}}$ to 
determine the values of $\lambda_{1},\lambda_{2},\lambda_{3}$ using equation 
(\ref{eqn:lam_hatdelc}) for each host halo, and finally reconstruct the 
axis-ratios of the host halo using equation (\ref{eqn:axis_lambda}). 
With the above method, we finally reconstruct the two axis-ratios of all 
four clusters at four redshifts. 

\begin{figure}
\begin{center}
\plotone{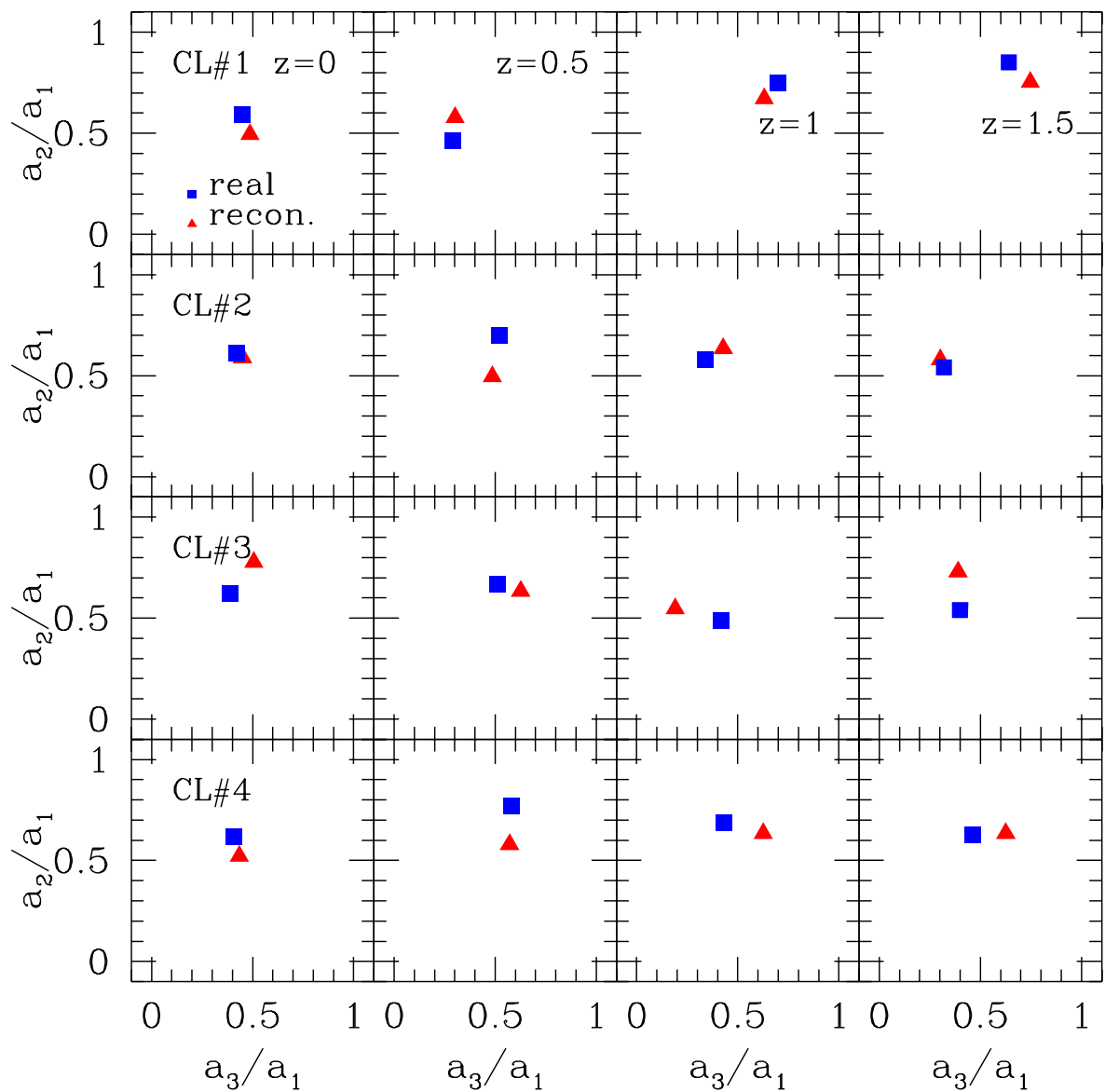}
\caption{Axis-ratios of four individual clusters at four different epochs; 
In each panel, the square represents the real axis-ratio measured in 
simulations while the triangle represents the reconstructed axis-ratio.
\label{fig:axis_ratio}}
\end{center}
\end{figure}

Figure \ref{fig:axis_ratio} plots the intermediate-to-major axis ratio vs. 
the minor-to-major axis ratios of the four clusters at four redshifts. 
In each panel, the (red) triangle dot represents the reconstructed 
axis-ratios by the above algorithm, and the (blue) square dot is the 
original axis-ratios from simulations. Although the reconstruction does 
not look perfect, it works quite satisfactorily, especially given all 
the approximations used in the algorithm. It is obvious that  
at least one of the two axis-ratios is fairly accurately reconstructed. 
We note that for the case of $s \simeq -1$, the difference between $a_2/a_1$ 
and $a_{2}/a_{1}$ is relatively large, which trend is also reconstructed 
well by our algorithm.
 
\section{DISCUSSIONS AND CONCLUSIONS}

Our achievements are summarized as follows: (i) we construct an analytic model 
for the anisotropic spatial distribution of dark halo substructures; 
(ii) we confirm numerically that the dark halo substructures have a 
strong propensity to be located near the {\it major} axes of their 
host halos; (iii) we develop an algorithm to reconstruct the triaxial 
shapes of dark halos from the anisotropic spatial distribution of their 
substructures.

Successful as are our analytic model and the reconstruction algorithm 
based on it, it is worth noting a couple of its inconsistencies. 
First, as mentioned in \S 2, equation (\ref{eqn:alpha_dis}) 
holds only when the tidal shear field is smoothed on the substructure 
radii. We ignored the smoothing scale difference, which should cause 
some scatters in the measurement of the correlation parameter $s$. 
Second, we adopted the spherical collapse condition 
$\lambda_{1} + \lambda_{2} + \lambda_{3} = \delta_{c} \approx 1.686$ 
even though we consider triaxial dark matter halos. Definitely, it 
will be necessary to refine our analytic model further, accounting for 
these two inconsistencies. 

Another limitation of our model is that its underlying assumption 
has been numerically tested only for the case of the concordance 
cosmology. Equation (\ref{eqn:axis_lambda}) was shown to work 
only provided that the background cosmology is described by the 
concordance cosmology \citep{lee-etal05}.  
To find the dependence of the correlation parameter, $s$, on cosmology, 
and to apply the model to a broader range of cosmological models, it will 
be first necessary to make more general assumptions and to test the 
underlying assumptions against numerical results, which is definitely 
beyond the scope of this paper. 
 
Now that we proved that it can reconstruct {\it in principle} the 
axis-ratios of triaxial dark halos from the anisotropic spatial 
distribution of substructures for the concordance $\Lambda$CDM model, 
it is worth discussing about how to apply the algorithm to real data 
from observations of such as the weak gravitational lensing.  
In our analytic model, the substructure positions are defined in the 
real three dimensional space relative to the principal axes of host halos, 
which is, however, very difficult to determine in practice.  
In observations they are measured in the two dimensional space projected 
along the line-of-sight direction which is not necessarily coincident 
with the host halo principal axis. Therefore, to apply our reconstruction 
algorithm to real data, it will be necessary first to find an analytic 
expression of the substructure spatial distributions in the general 
non-principal axis frame and to account for the effect of projection 
along the line-of-sight.  Our future work is in this direction, and 
we wish to report it elsewhere soon.

Another direction is to explain the Holmberg effect. 
As shown, our analytic model for the substructure spatial distribution 
(eq.[\ref{eqn:alpha_dis}]) is characterized by the single correlation 
parameter, $s$. By investigating what physical mechanism caused the 
correlation parameter to change their signs, it might be 
possible to explain the Holmberg effect theoretically.
 
Finally, we conclude that our analytic model may provide a physical 
quantitative understanding of the anisotropic spatial distribution of 
dark halo substructures, and will allow us to determine in principle 
the axis-ratios of triaxial dark halos from the observables.

\acknowledgments

We thank Y.Jing for generously providing simulation data and many useful 
discussions. We also thank an anonymous referee for careful reading and 
many helpful suggestions which helped us improve the original manuscript.
J.L. is supported by the research grant No. R01-2005-000-10610-0
from the Basic Research Program of the Korea Science and Engineering
Foundation.  X. K. is partly supported by NKBRSF (G19990754), by NSFC 
(Nos. 10125314, 10373012), and by Shanghai Key Projects in Basic Research 
(No. 04jc14079)

\clearpage

\clearpage

\end{document}